\documentclass[sigconf]{acmart}

\usepackage{booktabs} 

\setcopyright{rightsretained}

\copyrightyear{2017}
\acmYear{2017}
\setcopyright{acmcopyright} \acmConference{SafeConfig'17}{November 3, 2017}{Dallas, TX, USA}\acmPrice{15.00}\acmDOI{10.1145/3140368.3140372} \acmISBN{978-1-4503-5203-1/17/11}

\begin{document}
\title{Towards Realistic Threat Modeling: Attack Commodification, Irrelevant Vulnerabilities, and Unrealistic Assumptions}

\author{Luca Allodi}
\affiliation{%
  \institution{Eindhoven University of Technology}
  \streetaddress{Eindhoven, the Netherlands}
}
\email{l.allodi@tue.nl}

\author{Sandro Etalle}
\affiliation{%
  \institution{Eindhoven University of Technology}
  \streetaddress{Eindhoven, the Netherlands}
}
\email{s.etalle@tue.nl}

\renewcommand{\shortauthors}{L. Allodi and S. Etalle}

\begin{abstract}
Current threat models typically consider all possible ways an attacker can penetrate a system
and assign probabilities to each path according to some metric (e.g. time-to-compromise).
In this paper we discuss how this view hinders the realness of both technical (e.g. attack graphs) and strategic (e.g. game theory) approaches of current threat modeling, and propose to steer away by looking more carefully at attack characteristics and attacker environment. We use a toy threat model for ICS attacks to show how a realistic view of attack instances can emerge from a simple analysis of attack phases and attacker limitations.
\end{abstract}

%
%



\maketitle

\section{Threat modeling: introduction and background}

The importance of threat modelling in the assessment of one's security posture is
underlined by the numerous academic~\cite{Wang-2008-DAS,MANA-WING-11-TSE,Dolev-1983-IT,roy2010survey}, commercial~\cite{ibm2017security}, and standardized approaches~\cite{stoneburner2002sp,brenner2007iso,fredriksen2002coras} reported in the literature. In general, two broad categories
for threat modelling can be identified in the literature:

\paragraph{1) Vulnerability and technical exposure} System vulnerabilities can be exploited
by attackers to obtain unauthorized access to a system. The notion of `attack surface'~\cite{MANA-WING-11-TSE} reproduces the entry points of the attacker to then compute a metric of `exposure' to
attacks. The paths that an attacker can follow to reach a specific system or resource can be modeled through `attack graphs', where each node is a vulnerability on a system,
and edges represent the attacker's possibility to `escalate' to the next system~\cite{Wang-2008-DAS}. These graph often tend to explode as network size grows, making the identification of realistic paths to compromise difficult to achieve~\cite{obes2013attack}. In all approaches, the vulnerability plays a central role in the threat modeling: for example,
the probability of traversing an attack graph alongside a certain path is typically computed
as a function of some measurement of each vulnerability.

\paragraph{2) Strategic approaches} A different perspective is adopted by models that account for 
attacker strategies, often in response to a defensive posture by their target~\cite{roy2010survey}. These models are typically specified using classic game-theory, and solved looking for specific equilibria: for example, Nash solutions assume that each player (attacker and defender) aims at maximizing his/her own outcome irrespective of the choice of the other player (accounting instead for the possible decisions he/she may make); Stackelberg security games assume instead that the attacker's decision depends on the observation of the defender's strategy. The model solutions oftentimes assume that `all attackers look alike'~\cite{abbasi2016know} or that players act over perfect information~\cite{fudenberg1991game}. These assumptions also underlie recent linear-programming solutions to the defender's problem~\cite{Serra-TISSEC-2015},
where the attacker is implicitly assumed to be all-knowing with respect to the vulnerabilities on the system.


Despite the diversity in the adopted perspective or application scenario, all current approaches have an underlying assumption in common: \emph{the attacker can arbitrarily choose whichever attack vector or sequence he/she thinks will maximize the function the model assigns to him or her}. For example, attack graphs enumerate all possible (known) vulnerabilities that the attacker may exploit; game theoretic approaches assume the attacker can choose whichever attack strategy he/she believes best, irrespective of the solution strategy; NIST 800-30 directives require the assessors to look at all possible attack paths toward an asset. This of course puts the defender at high strain as he/she needs to defend against \emph{all} vulnerabilities (ranging in the thousands~\cite{NVD}), whereas the attacker only need one or a few, and can choose any.

\section{Empirical views on attacks}




Recent empirical findings, replicated independently over a number of studies, present 
results in sharp contrast with the general assumption that attackers may exploit any available vulnerability. Whereas from this threat model one would expect exploits to appear in attack signatures and network telemetries at comparable rates (i.e. ideally, uniformly), the empirical observation is \emph{widely} different: out of tens of thousands of possible vulnerabilities only a fraction is actively exploited in the wild (even after controlling for observational biases)~\cite{Allodi-2014-TISSEC}. This effect is even stronger when looking at the actual volumes of attacks driven by each (exploited) vulnerability, for which we observe heavy tail distributions of attacks~\cite{Nayak-2014-RAID,Allodi-ESSOS-15}. This effect also emerges for so-called 0-day vulnerabilities, i.e. vulnerabilities known to attackers before vendors have an opportunity to fix them~\cite{Bilge-12-CCS}: over an historical record of attacks worldwide, Bilge and Dumitras find only less than twenty 0-days, of which two cumulatively carry two million attacks, and the remaining constitute negligible background noise five orders of magnitude lower.

The figures remain similar by looking at the exploits attackers trade in the black markets: exploit kits drive millions of attacks~\cite{Grier-12-CCS,Provos-2008-USENIX,symantec-ekits-2011}, yet they embed five to ten exploits each at a maximum~\cite{Kotov-2013-ESSOS}, a figure settled around 3-4 exploits per kit in recent years~\cite{allodi2017ccs,ablon2014markets}. At the same time, the vast majority of attacks in the wild are carried by this handful of vulnerabilities only~\cite{Allodi-2014-TISSEC,Allodi-13-IWCC}, and the refresh time of attacks in the wild is as slow at 600 days (i.e. the same exploits are re-used in the wild for almost two years before they are substituted at scale with a new attack)~\cite{Allodi-17-WAAM}.
Similarly, attacks against Industrial Control Systems like SCADA, Distributed Control Systems, and PLCs, that form the skeleton of the critical infrastructure, seem to be relatively rare. While the severity of these attacks, especially when perpetrated at a national or infrastructural level, is certainly high, ICS incidents seem to affect a very limited fraction of the overall vulnerable infrastructure~\cite{ibm2017security,cisco2017annual}.

The present empirical view on attack distributions can not, clearly, 
justify the assumption that attackers will choose any vulnerability among those available to deliver their attack. Even if systems remained vulnerable at non-uniform rates (which appears not to be the case~\cite{nappa2015attack}), one would expect the overall distribution of attacks to 
`spike' for vulnerabilities preferred by attackers (e.g. those that get patched at a slower rate
than the average), \emph{not} for all vulnerabilities to be collapsed several orders of magnitude below the preferred set, a phenomenon all available data clearly shows~\cite{allodi2017ares}.
This is clearly unacceptable because  the defender needs to decide which vulnerabilities to fix first, and it becomes therefore critical to identify which vulnerabilities are most likely to drive orders of magnitude more attacks than the rest.

\section{Attack generation}


It remains therefore an open question to identify the `attack generation process' that
out of thousands of vulnerabilities selects only a small fraction that will deliver the vast majority of attacks. This is reflected in recent work, carried independently by other authors, questioning the
soundness of current attacker assumptions~\cite{kelly2017bh}, in particular in the `hunting' of game-theoretic solutions that attackers operate under complete information, and act rationally
in their decision process. In related work in the cognitive sciences, Veksler  showed that, depending on the payoff matrix, game theoretic solutions to the attacker's mixed strategy may lead the attacker to lose in 50\% of
the scenarios, i.e. the attacker is indifferent to the choice and randomly
picks an attack strategy~\cite{Veksler2016KnowYE}. On this same line, Abbasi et al. conclude that a `homogeneous' distribution of attackers can not be assumed when solving for equilibrium solutions to a game theoretic model~\cite{abbasi2016know}.
Overall, one can broadly distinguish between `commodified' and `tailored' attacks as threat sources:

1) Commodified attacks are generated by some pre-existing attack mechanism that is available to the attacker through the environment in which he or she operates. These report the vast majority of attacks~\cite{Provos-2008-USENIX,Nayak-2014-RAID,Allodi-13-IWCC}. For example, an attacker may buy an exploit kit from the underground cybercrime markets~\cite{Grier-12-CCS,Allodi-TETCS-15,allodi2017ccs}, or a member of the self-proclaimed hacktivist organization \emph{Anonymous} may largely use a pre-engineered tool to launch a distributed denial of service attack. Importantly, the update rate of these attacks is not realistically driven by the state of any specific target, but rather by the state of the \emph{population} of
possible targets: for example, an exploit kit gets realistically updated when the overall efficacy of the exploit it embeds drops below a certain (commercially viable) threshold~\cite{Allodi-17-WAAM} (corresponding to the well-known economic process of `inaction' first formalized by Stokey in 2008~\cite{stokey2008economics});

2) Tailored attacks are attacks that are carried against a specific target or a specific set of targets. These require some level of engineering or technical sophistication on top of the mere deployment of the attack on the side of the attacker. For example an attacker may need to modify an exploit to fit the specific environment of the target system, or engineer a new attack altogether after finding that none of the attacks in his/her possession matches an existing vulnerability~\cite{allodi2017RA}. Importantly, in these scenarios the attacker
often acts under severe uncertainty in the topology, structure, and characteristics of the attacked network, which makes the identification of a `one-size-fit-all' attack strategy impossible to achieve~\cite{sarraute2012pomdps}.

The former class of attacks has been recently covered in a number of papers highlighting 
the mismatch between empirical observations and attacker assumptions~\cite{kelly2017bh,Allodi-17-WAAM,Nayak-2014-RAID,allodi2017RA}. 
`Tailored' attacks are however less investigated, on the one hand because much less common than `commodified' attacks~\cite{ten2008vulnerability}, and on the other because a clear
rationalization of attacker characteristics remains uncertain~\cite{grossklags2010price}.
For example, multi-stage attacks where the attacker needs to overcome a set of perimetral defenses to breach a network and then target an attack once the `internal' systems are revealed may require significant efforts in terms of reconnaissance and exploit engineering. This process may take several weeks or months to complete: whereas time-to-compromise is often proposed as a metric to evaluate `likelihood' of attack~\cite{Holm-TDSC-2014,nzoukou2013unified}, it remains unclear at what level of the attacker strategy this has an effect. For example, a targeted (`tailored') attack against a governmental facility perpetrated by political activists requires a specific timing to maximize media coverage (e.g. close to the approval of a controversial law); similarly, espionage requires an attack to be completed before the target secret loses tangible value for the attacker (e.g. classified information affecting stock value of an organization), or a small criminal organization may not have the resources to dedicate months in finalizing an attack against an ICS system.

These effects should be accounted for in the same fashion K. Shortridge~\cite{kelly2017bh} covers in her work: by considering the attacker's ability to learn during the attack, the set
of information he/she possesses before launching the attack, and his/her risk aversion to, for example, failing the attack (much like adversarial modeling in terrorism contexts prescribes~\cite{ezell2010probabilistic}).

\subsection{An example on a typical ICS system}

\begin{figure}
\includegraphics[width=0.85\columnwidth]{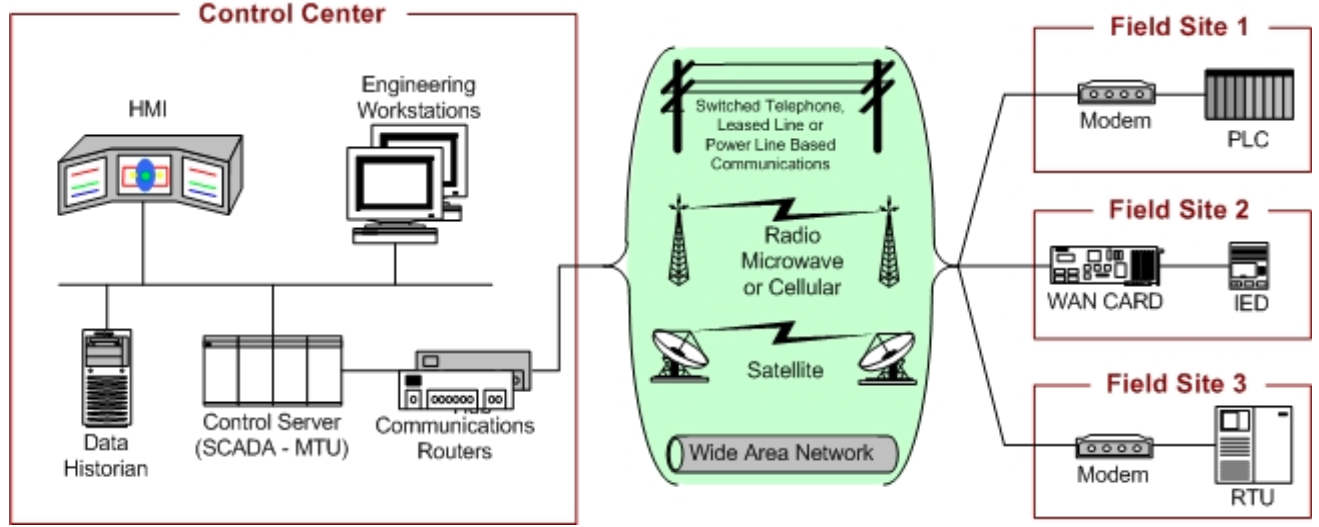}
\caption{Typical two-level ICS configuration. Picture taken from NIST Special Publication 800-82~\cite{stouffer2011guide}.}
\label{fig:ics}
\end{figure}

Figure~\ref{fig:ics}, drawn from NIST Special Publication 800-82 on ICS security~\cite{stouffer2011guide}, shows a typical multi-level ICS network whereby a control center communicates over one or many network layers with field sites where PLCs, specialized controllers, and sensors are typically located. In this structure, the systems in the control
center are typically `regular Windows systems'\cite{shodanreport} enabling SCADA control functionalities; these systems are ideally `air-gapped'~\cite{stouffer2011guide} (i.e. do not have any physical connection with other networks), but in practice they are often exposed to the open Internet to ease network management; the operation network is in the best case a VPN connecting the control systems to the peripheral
devices. The systems in the field sites are ultimately the real target of an attack as they are the ones that cause the impact in the physical world (e.g. accelerating a turbine, shutting off a power generator, etc.)~\cite{ten2008vulnerability}.

Following indications from the ICS-CERT,\footnote{\url{https://ics-cert.us-cert.gov/content/overview-cyber-vulnerabilities}.} this setup forces the attacker to structure his attack in three stages, as follows: (1) Penetrate the first  control center layer; (2) Analyze the internals of the target system; (3) Strike the target system(s).
 
A typical attack scenario involves a watering hole attack, or a targeted phishing email, to penetrate the first layers of the control center. In this stage the attacker can use a `commodified' set of attacks to gain control over what is, from a vulnerability perspective, a regular (hopefully hardened) Windows system. The actual control systems may require additional effort to be reached once inside (e.g. because of network segregation)~\cite{SANS2016blackenergy}. 
From here, extensive levels of reconnaissance are typically needed to (1) identify the network and system configuration visible from the first attack stage, and (2) understand which attacks are needed to move to the next phase.
For example, Kim Zetter's coverage of the Ukrainian Power grid attack~\cite{SANS2016blackenergy} summarizes:\footnote{\url{https://www.wired.com/2016/03/inside-cunning-unprecedented-hack-ukraines-power-grid/}}
\begin{quote}
The initial intrusion got the attackers only as far as the corporate networks. But they still had to get to the SCADA networks that controlled the grid [..] so the attackers were left with two options: either find vulnerabilities that would let them punch through the firewalls or find another way to get in. [..]
Over many months they conducted extensive reconnaissance, exploring and mapping the networks and getting access to the Windows Domain Controllers [..] Once they got into the SCADA networks, they slowly set the stage for their attack.
\end{quote}

\subsubsection{Toy model of a `tailored' attack}

Let us then consider an attacker that values the potential overall havoc that can be inflicted to the SCADA infrastructure as $V$. A certain fraction $\alpha$ of $V$ can be extracted from the first-phase attack (for example by spreading ransomware or by defacing a public server). As the first stage can be aided by commodified attacks, we assume the attacker succeeds with probability 1~\cite{Kotov-2013-ESSOS,Allodi-13-CSET}. The payoff for the first phase is:
\begin{equation}
\label{eq:phaseOne}
\Pi_1 = \alpha V - C_1
\end{equation}
where $C_1$ is the cost of performing the attack (e.g. to buy the exploit).

From the newly acquired position the attacker may start performing some level of reconnaissance on the internal network (stage 2), including the identification of other subnets, bridge systems to the SCADA control network, and/or the topology of the ICS network (see Fig.~\ref{fig:ics}). Let us denote with $p_2$ the overall probability that the attacker successfully accomplishes this phase. Then, with probability $p_2$ the attacker extracts the remaining fraction of $V$, i.e. $(1-\alpha)V$ and pays an upfront cost $C_2$ for completing phase-two. As discussed above, an attacker may wish to finalize the attack within a specific timeframe, e.g. because of limited resources to commit for prolonged attacks. The overall value of phase-two is discounted by a factor $e^{-\delta t}$.
We therefore set the two-phase attack profit as:
\begin{equation}
\label{eq:phaseTwo}
\Pi_2 = \alpha V - C_1 + p_2\left[(1-\alpha) V \right]e^{-\delta t} - C_2
\end{equation}

\begin{figure}
\centering
\includegraphics[width=0.85\columnwidth]{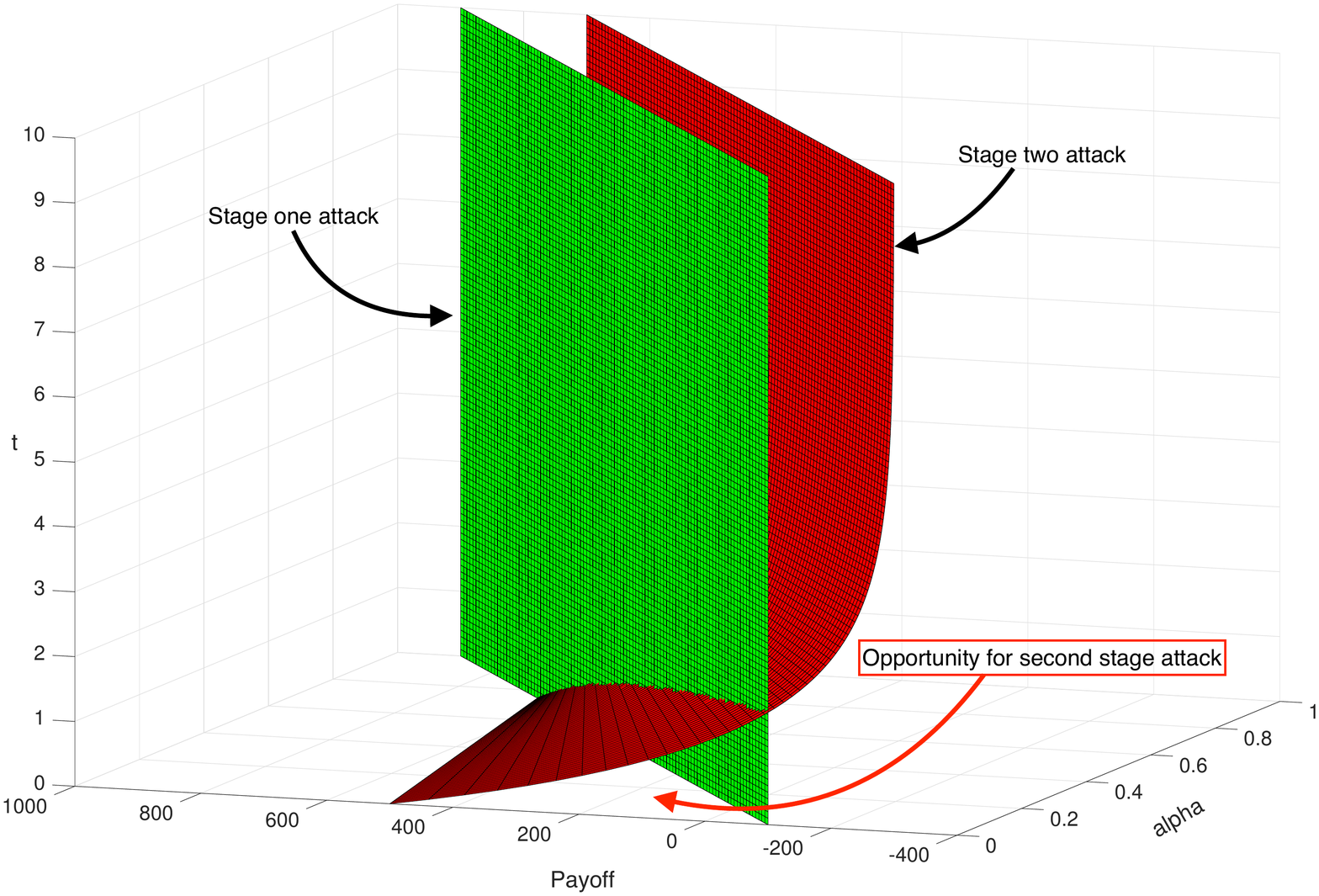}
\caption{Numerical simulation of phase-one and phase-two attacks. $p=0.8, V=1000, \delta=0.8, C_1=0.1\times V, C_2=2\times C_1$.}
\label{fig:sim1}
\end{figure}

Figure~\ref{fig:sim1}
reports a numerical simulation of the two models. Negative payoffs indicate areas where the attacker will not act. As one can see, the value of a two-phase attack quickly decays as $T$ increases (as $\lim_{t\to\infty} \Pi_2 = \Pi_1-C_2$), and the rate of decrease increases as $\alpha$ grows. 
This reflects the intuition that a two-phase attack quickly loses of appeal as time passes, and as the value that can be extracted from a phase-one attack increases.

\subsubsection{Take-aways}

From the example analysis above it emerges that attackers in the second phase will only materialize 
in the indicated area under the curve in Fig.~\ref{fig:sim1}: for example, hacktivists and other low-resource attackers will largely disregard phase-two attacks (which quickly lose of appeal as too many resources and time likely need be dedicated to complete the second phase), and focus instead on standard internet-facing systems, which appears indeed to be the case for most attacks of this type~\cite{verizon2016data,bronk2013cyber}. For example, infecting an organization's network with ransomware may be enough to cause the desired level of havoc, or a website defacement enough to reach the desired audience (e.g. the organization's customers)~\cite{verizon2016data}. The appeal of stage two attacks decreases the lower their overall impact on the targeted infrastructure is (i.e. as $\alpha\rightarrow 1$). Differently, nation-state or powerful attackers may have the adequate resource set to endeavour in the second-stage attack, i.e. because they have a set of ready-to-use SCADA exploits, or they have a discount factor close to zero (i.e. they have resources to commit to the attack for long periods of time). 

Under this model, the appearance of a new vulnerability in an ICS system carries little weight
on the overall risk scenario, as most attackers would not be able to exploit it (unless commodified). Reasoning about the attacker's ability to learn at every stage of the attacker can help in setting parameters of the model to realistic values and, critically, in understanding the correct shape of the learning function of the attacker. The presence of multiple attackers,
or collusion of attackers with insiders, may be considered in the model as well, in the same fashion as terrorist cells are modeled in the risk analysis literature~\cite{ezell2010probabilistic}. 

Importantly, in this model we are assuming that the value $V$ of the attacked infrastructure is determined exogenously; in reality, attackers may adjust $V$ depending on their own specific interest. In this simple model, $\alpha$ represents the mechanism that associates the value assigned by the attacker to the first stage attack. For example, a human rights hacktivist may be reluctant to exfiltrate private costumer data, and assign a high value to the defacement of public services of the target; on the other hand, a market competitor  may have the opposite view. In reality, attacker payoffs may be influenced by several endogenous factors that determine his or her true preference; for example, an attacker may value \emph{success} of attack
more than \emph{impact} of attack, i.e. prefers to select a strategy that yields the highest chances of success despite low consequences over a strategy with low probability of success but that delivers catastrophic impact~\cite{ezell2010probabilistic}. These considerations are often reported in the \emph{risk analysis} literature, but seldom considered in the cybersecurity field, where the `technical' perspective oftentimes overcomes considerations on attacker skills, goals, and capabilities.

\bibliographystyle{ACM-Reference-Format}
\bibliography{short-names,security-common,bib,SCADA_encryption,SCADA_statistics} 

\end{document}